\documentclass[10pt,a4paper]{article}
\usepackage{amsmath}
\usepackage{verbatim}
\usepackage{graphicx}
\usepackage{natbib}
\usepackage{pslatex}
\usepackage{hyperref}
\bibpunct[; ]{(}{)}{;}{a}{}{;}

\newcommand{\req}[1]{(\ref{#1})}
\newcommand{\be}{\begin{equation}}
\newcommand{\ee}{\end{equation}}
\newcommand{\bea}{\begin{eqnarray}}
\newcommand{\eea}{\end{eqnarray}}

\newcommand{\avg}[1]{\langle{#1}\rangle}

\newcommand{\BE}{\begin{eqnarray}}
\newcommand{\EE}{\end{eqnarray}}
\newcommand{\BEn}{\begin{eqnarray*}}
\newcommand{\EEn}{\end{eqnarray*}}
\newcommand{\barr}{\begin{array}}
\newcommand{\earr}{\end{array}}

\newcommand{\bit}{\begin{itemize}}
\newcommand{\eit}{\end{itemize}}
\newcommand{\bc}{\begin{center}}
\newcommand{\ec}{\end{center}}
\newcommand{\ben}{\begin{enumerate}}
\newcommand{\een}{\end{enumerate}}

\begin{document}

\title{The universal shape of economic recession\\ and recovery after a shock\thanks{We thank Vladimir Popov and Paul Ormerod for their public comments, and David Br\'ee for his critical reading of the first version of this manuscript. This work has been supported in part by the EU projects DAPHNet (ICT grant 018474-2) and CO3 (NEST grant :12410)
}}
\author{Damien Challet$^{1,2}$, Sorin Solomon$^{3,2}$ and Gur Yaari$^{3,2}$\\$^1$ D\'epartement de Physique, Universit\'e  de Fribourg, P\'erolles, 1700 Fribourg, Switzerland\\$^2$ Institute for Scientific Interchange, via S. Severo 65, 10113 Turin, Italy\\$^3$ The Racah Institute of Physics, Hebrew University, Jerusalem,  91905, Israel}

\maketitle

\begin{abstract}
We show that a simple and intuitive three-parameter equation fits remarkably well the evolution of the gross domestic product (GDP) in current and constant dollars of many countries during times of recession and recovery. We then argue that this equation is the response function of the economy to isolated shocks, hence that it can be used to detect large and small shocks, including those which do not lead to a recession; we also discuss its predictive power. Finally, a two-sector toy model of recession and recovery illustrates how the severity and length of recession depends on the dynamics of transfer rate between the growing and failing parts of the economy.\\

\noindent keywords: Economic growth, GDP, shocks, response function,  modelling, prediction, optimal policy\\JEL: C32, O23, O41
\end{abstract}

\newpage
\section{Introduction}

Explaining growth and recessions has been central to Economics ever since its beginning \citep{deQuesnay,AdamSmith,Ricardo,Keynes,Solow,Schumpeter,Romer,HistoryEconomyLSE}. Since recessions and subsequent recoveries are usually split into distinct episodes in economic analysis, the factors of decline and growth have been investigated separately (e.g. \cite{Popov2006,Kolodko,ULJ}) and little attention has been devoted to the intrinsic relationship between recession and recovery.

Here we argue that
\begin{enumerate}
\item recessions and their subsequent recoveries can be fitted rather well by a single 3-parameter function that contains both the recession and the recovery parts. It assumes that during recession-recovery periods, at any time a fraction of the economy is shrinking exponentially while the rest is growing exponentially. As a consequence the two parts of the GDP curve are intrinsically linked and cannot be considered as separate events. In particular, we show why this yields much better estimates of the decaying and expansion rates.

\item  The shape of this function is the {\em simplest one} that respects the underlying economic process: economic activity grows and shrinks exponentially. A more complex superposition of exponentials or non-constant parameters is of course possible, as discussed below.

\item It is valid as long as no other shock occurs, thus can be used {\em a contrario} to separate a GDP time series into episodes of economic growth, providing a factinating new way of reading the fluctuations of GDP time series, even outside times of recessions. This leads to the conclusion that this model is in fact the {\em response function} of the economy as a whole to rare negative shocks, both exo- and endogeneous.
\end{enumerate}


In section \ref{sec:data}, we consider the yearly evolution of countries having experienced lasting recessions others than those due to wars, which include many of the former communist block economies following their liberalisation. We find that our ``universal recession-recovery shape'' fits all of them well {\em between shocks} and that each additional shock brings a new episode fitted by our equation.  Its accuracy, and in particular, the smooth shape of GDP evolution between shocks it implies, is confirmed to a high degree by quarterly data for Finland and the United Kingdom. We illustrate how this equation can be used to detect additional shocks automatically.

In section \ref{sec:th}, as a theoretical exercise, we introduce a simple two-sector model of growth with economic transfer. We also discuss why a simple two sector model is able to reproduce faithfully the typical global dynamics of recession-recovery. Then we exploit this understanding in order to find an effective transfer rate that minimises the depth and duration of the recession, and maximises both the GDP value and the final growth rate. We finally propose a means to differentiate static from dynamical effective policies in historical data.

\section{Data analysis}
\begin{figure}\centerline{\includegraphics*[width=0.7\textwidth,angle=270]{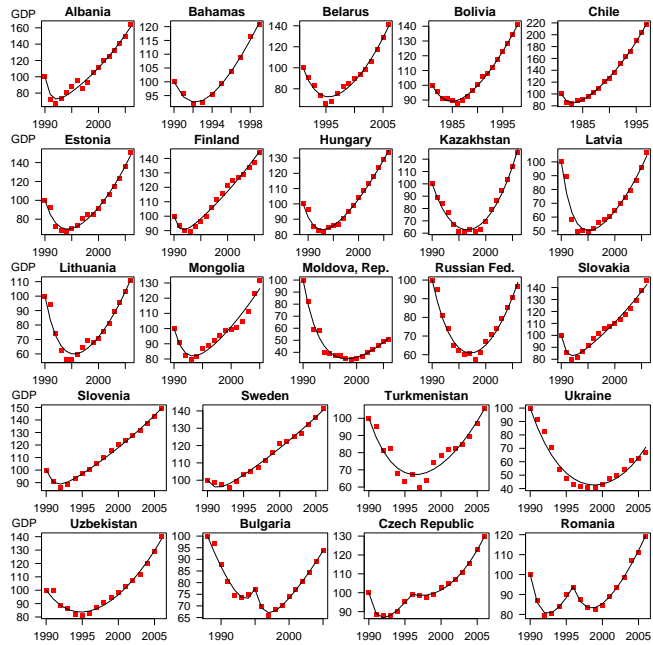}}
\caption{Time evolution of the GDP in current dollars of 23 countries, and the fits to Eq.\ \req{eq:g(t)}}
\label{fig:gdp_current}
\end{figure}

\begin{figure}
\centerline{\includegraphics*[width=0.7\textwidth,angle=270]{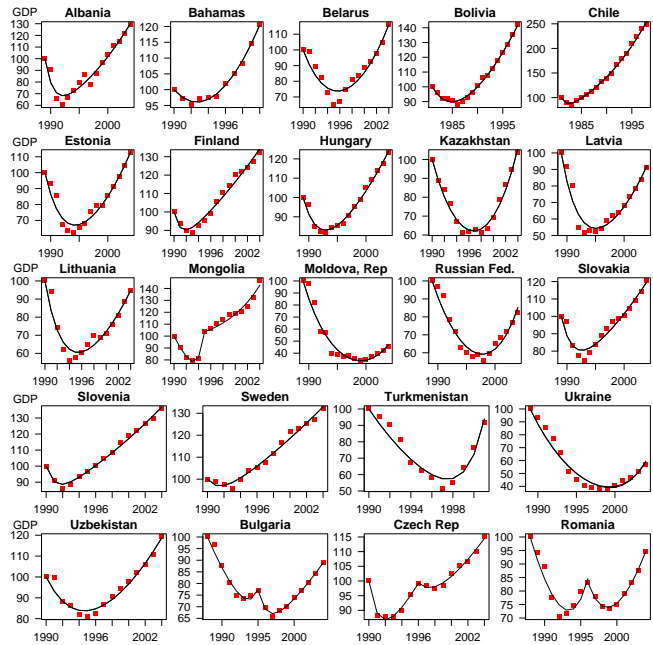}}
\caption{Time evolution of the GDP in constant dollars of 23 countries and the fits to Eq.\ \req{eq:g(t)}}
\label{fig:gdp_constant}
\end{figure}

\label{sec:data}
The economic activity of many countries shows dramatic and sudden decreases followed by a slow road to recovery, a pattern commonly known in  Economics as L, U or J-shape depending on the state of unfolding of the J shape \citep{ULJ}. Our dataset, taken from \url{http://data.uno.org}, starts in 1980. We retained the 23 countries having undergone recessions of at least three consecutive years, as we need enough points in the recession part to estimate correctly the rate of decay. Among these countries one finds many of the previously communist Eastern European countries; Finland, because of the strong links between its economy and those of the Eastern Block, suffered from the same crisis; Sweden had a crisis of its own at about the same time, caused by the burst of a real estate and financial bubble. Bolivia, Bahamas and Chile follow the same pattern, at different times, for reasons unknown to us.

\subsection{Yearly GDP data}

We focus on gross domestic product (GDP) both in current and constant dollars. The latter is also called real GDP as it discounts inflation and makes is possible to better compare economic activity from year to year and between countries. As we shall see, however, it is more irregular, in part because of the dynamics of currency exchange rates, which in turn reflects partially that of interest rates and differences in budget deficits.

As it turns out, most of these recessions were caused by political and taxation reforms or financial crises; in other words, by events localized in time, that is, sudden shocks. In order to cause long recessions, these shocks must be large and affect much of the economy. One key assumption in the following is that the intrinsic economic parameters (fraction of the economy affected,  growth and decay rates) are or appear to be constant ever after a shock. We shall not attempt to model the occurrence and properties of shocks. Figure \ref{fig:gdp_current} reports the evolution of the GDP in current dollars of several countries,  revealing a common pattern. These recessions can be characterised by their intensity (maximum loss of GDP), the time of the minimum GDP and the time to recover the previous level of economic production. The evolution of the GDP in constant dollars (Fig.\ \ref{fig:gdp_constant}) shows similar patterns.

Remarkably, all previously communist Eastern European and Soviet Union countries have experienced a lasting recession followed by a recovery. Some countries such as Romania, Bulgaria and the Czechia clearly display a double dip (Figs \ref{fig:gdp_current} and \ref{fig:gdp_constant}). As it happens, the onsets of the second dips of Romania and Bulgaria is unambiguously related to a change of power, suggesting that wrong economic policies or implementations may  be blamed for further degradation of the situation.

The severity and duration of the recession varies widely between the countries: Poland has recovered very quickly (too much so to be shown in Figs.\ 1 and 2), while some countries such a Russia and Latvia have come back to their previous current-dollar GDP in 2006. The Republic of Moldova was still 50\% down in 2006. We shall not try to find the causes of such differences, although our fitting equation directly provides information about the fraction of the economy which is decaying, but shall discuss various ways of optimising a policy with the help of a simple theoretical model in secion \ref{sec:th}.

For reasons explained below, we parametrise the GDP J shapes as displayed in Figs\ \ref{fig:gdp_current} and \ref{fig:gdp_constant}, in terms of the following formula
\be\label{eq:g(t)}
W(t)=W(t_0)[f e^{\lambda_+(t-t_0)}+(1-f)e^{\lambda_-(t-t_0)}],
\ee
 where
\bit
\item $W(t_0)$ is the initial GDP at the time $t_0$ of the reform
\item $f$ is the fraction of the economy that grows at rate $\lambda_+$;
\item the rest of the economy ($1-f$) deflates at rate $\lambda_-$
\eit
A J-shaped $W$ is obtained if $W'(t_0)<0$, i.e. if $f\lambda_++(1-f)\lambda_-<0$.

The fit of recession and recovery times of 23 countries appears remarkable (the details are reported in appendix A). Since it uses constant parameters, it may a priori suggest a surprising degree of constancy of rates of decline and growth; however, as discussed in section 4, dynamic effective  policies can also be relatively well fitted with the same model: one needs data more detailled than the global GDP time series to detect them.

The fitting function of Eq.\ \ref{eq:g(t)} can be considered as the simplest model of recession and recovery that is economically meaningful. Indeed, it only assumes that one part of the economy shrinks while the other one grows, both exponentially. It does not include many other {\em a priori} relevant parameters. It might seem tempting to fit the dips with a parabola, which of course give terrible results as the GDP curves are not symmetric and have an asymptotic constant exponential growth rate. Because of the auto-catalytic nature of economic growth and recessions, only a sum of exponential makes sense, which also excludes other candidates such as splines (which need many more than 3 parameters).

Given the simplicity of the fitting function and the complexity of the underlying process, it is to be expected that the fits should not be perfect. Beyond gentle noise,  one finds two types of deviations. First, during the recession phase, the GDP of some countries (e.g. Albania, Estonia, Hungary, Latvia, Turkmenistan) is concave, especially in the very first part of the recession, which is impossible to achieve with the above equation. This probably comes from the fact that our model implicitely assumes that the economic shock affects all the sectors of the economy at once. A solution would be that the fraction of the deflating part $1-f(t)$ increases, for instance $1-f(t)=(1-f)[1-e^{-t/\tau}]$ where $\tau$ is the typical speed at which the shock propagates throughout the economy.

Another story is told by systematic irregularities in the recovery part: most of them are negative deviations. Many countries experienced a temporary pause in their growth, sometimes a short-lived recession, and a few countries a lasting secondary recession. Because we have only annual data, and since one needs at least three consecutive points for a good fit of the recession part, we chose in some cases to fit the whole time series instead of fitting these episodes separately. The use of quarterly data solves this problem and, importantly, allows one to uncover additional economic shocks that do not necessarily lead to recessions, as seen in subsection \ref{subsec:Finland}.  We could trace possible causes of some additional shocks: Albania suffered from a bank crisis in 1996, and Finland from the burst of the 2000 internet bubble (see also subsection \ref{subsec:Finland}). The worst secondary shocks were born by Romania and Bulgaria (1996 elections), and Czechia (2000 bank crisis) which resulted into a second dip. This raises the question on whether the second reforms were successful or on the contrary detrimental: comparing the values of $\lambda_+$, $\lambda_-$ and $f$  before and after the secondary shock, one concludes that, according to our model, the crisis in Czechia had long-term  positive effects, both the rates of expansion and decline  having much improved, at the cost of the initial fraction of the expanding sector. The case of Romania is best described as {\em bis repetita (non placuerunt)}: the second crisis leading to almost the same fitting parameters as the first one. Finally, Bulgaria has spurious results regarding the first shock (GDP in constant dollars), which is due to the fact that GDP was mostly decreasing, hence the fit could not assess correctly the growing part of the economy, finding it very small (from $0.2\%$ to $2\%$), but doing very well. This result is clearly wrong but easily reproduceable with other countries if one restricts the data so as to only include a very small part of the recovery. The figures obtained for the second shock are in line with all the other shocks experienced by the other 22 countries.

\subsection{Quarterly data}\label{sec:quarter}

\begin{figure}
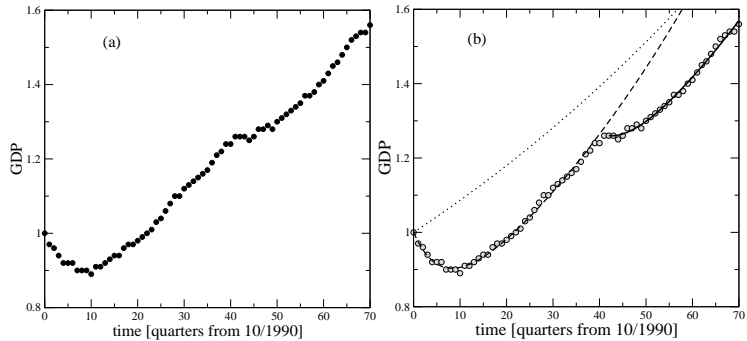

 \centerline{\includegraphics*[width=0.4\textwidth]{fin.eps}\includegraphics*[width=0.4\textwidth]{fin_fits.eps}}
 \caption{Scaled real quarterly GDP of Finland from 04/1990 without fit (left) and with two fits and the continuation of the pre-1990 growth trend (dotted line).}
\label{fig:fin}
\end{figure}

Quarterly data is available for the recent history of industrialised countries. Let us first focus on Finland whose 1990 recession was analyzed in the previous subsection. Figure \ref{fig:fin}a plots the real GDP of Finland without any fit. While the recession and subsequent recovery are easy to spot with naked eyes, it is much harder to make sense of the fluctuations in the recovery part. As claimed above, the fitting equation that we propose provides a new way to interpret such plots.  Figure \ref{fig:fin}b adds the fit of the 1990 recession, which is even more impressive than for yearly data (note that we do not use lin-log axis, that would make the fit even more impressive). Given the faithfulness of our three-parameter model for the first 41 quarters (more than 8 years),  one must conclude that something sudden happened to the GDP dynamics. This can be traced  to the burst of the Internet bubble and subsequent weak growth of the world during these years, which resulted in a sharp decrease of demand for paper products and mobile phones, which contributed to half of the exports of the country at the time. Assuming that the shock was restricted in time, we can once again fit the next period as nicely with the same equation. It is revealing to compare the parameters of the fits of the three episodes (before 1990, 1990--2001, 2001-2008). We have $\lambda_{+}^{(-1990)}<\lambda_{+}^{(2001-2008)}<\lambda_{+}^{(1990-2001)}$: whereas the post-2001 GDP cannot catch up with the asymptotic 1990--2001 trend, there was hope before the current global recession that it would overtake the continuation of the pre--1990 trend.

\begin{center}
\begin{table}
\begin{tabular}{|c|c|c|c|}
 \hline
 Finland&$f$&$\lambda_+$&$\lambda_-$\\ \hline
\hspace{4ex}--1990        &NA&$0.0082\pm0.0001$&NA \\ \hline
 1990--2001 (0--42)& $0.768\pm0.006$&$0.0121\pm0.0002$& $-0.176\pm0.010$ \\ \hline
 2001--2008 (42--71)& $0.925\pm0.016$&$ 0.0106\pm 0.0006$&$ -0.143\pm0.040$ \\ \hline
 \end{tabular}
 \caption{Result of the fits of Eq.\ \req{eq:g(t)} to the real GDP of Finland.}
 \label{tabl:fin}
 \end{table}
 \end{center}

\begin{figure}
 \centerline{\includegraphics*[width=0.7\textwidth]{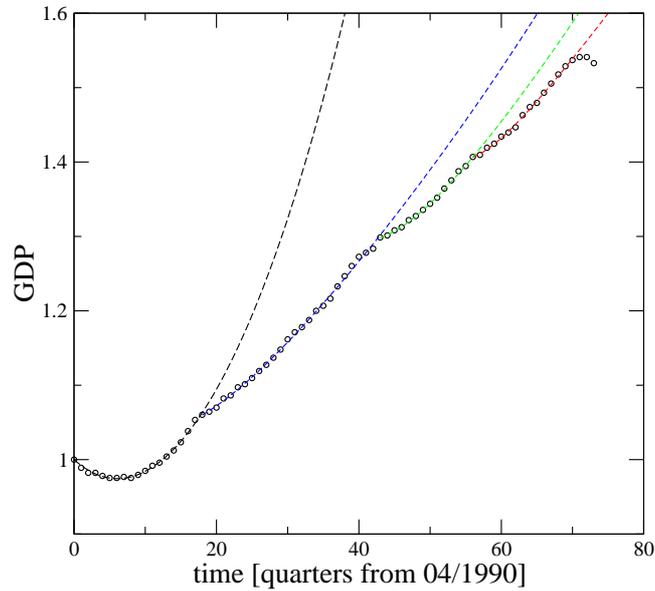}}
 \caption{Scaled real GDP of the United Kingdom with four fits corresponding to different episodes of economic growth.}
 \label{fig:uk}
\end{figure}

The analysis of the United Kingdom from 1990 to 2008 tells us the same story (see Fig.\ \ref{fig:uk}). There a first clear recession/recovery pattern beginning in 1990; the results of the fit are reported in table, which indicate that the growth rate was unsustainable; it is in fact related to a real estate bubble. After the latter broke up in 1995, two more shocks can be seen in 2001 and 2004. The fits of these three last parts give consistent results: their asymptotic growth rates  $\lambda_+=$0.0092, 0.0090, 0.0074 are comparable; they correspond to an annual growth of slighty more than 3\%.  The point is that none of these three shocks did lead to a recession, but all of them interrupted the trend of the GDP.  Each of these negative shocks are not large, but each of them is associated with a loss of absolute GDP in the asymptotic regime. The last shock of the time series corresponds of course to the ongoing recession.

\begin{center}
\begin{table}
\begin{tabular}{|c|c|c|c|}
 \hline
 UK&$f$&$\lambda_+$&$\lambda_-$\\ \hline
 1990--1995 (0--18)& $0.53\pm0.09$&$0.028\pm0.005$& $-0.049\pm0.010$\\ \hline
 1995--2001 (19--43)&
 $0.98\pm0.01$&$ 0.0092\pm0.0005 $&$ -0.151\pm0.106 $\\ \hline
 2001--2004 (43--55)&$0.96\pm0.02$&$0.0090\pm0.0012 $&$-0.154\pm0.073$\\ \hline
 2004--2008 (55--71)&$0.98\pm0.02$&$0.0074\pm0.0012$&$-0.168\pm0.241$  \\ \hline
 \end{tabular}
 \caption{Result of the fits of Eq.\ \req{eq:g(t)} to the real GDP of the United Kingdom.}
 \label{tabl:UK}
 \end{table}
 \end{center}

These two examples strongly suggest that economic shocks come in all sizes and that the function we propose is able to capture what happens to the GDP after them, hence the claim that Eq.\ \req{eq:g(t)} is the dynamical response function of the GDP.

\subsection{Making predictions}

Since this model fits well real data between two shocks, it must be able to make predictions assuming that no additional shock occurs, and by extension, should be able to detect an additional shock. Let us first start with noisy synthetic data in order to understand better its predictive power. Generating time series $G(t)=[f e^{\lambda_+t}+(1-f) e^{\lambda_-t}][1+\nu \eta(t)]$ where $\eta(t)$ are drawn at random from a unit-variance Gaussian distribution with zero average and $\nu$ is the strength of the noise. The issue is to determine the minimum length of data needed in order to determine faithfully the parameters, in particular if from a partial time series one can predict the future evolution of the GDP.

\begin{figure}
\centerline{\includegraphics*[width=0.7\textwidth]{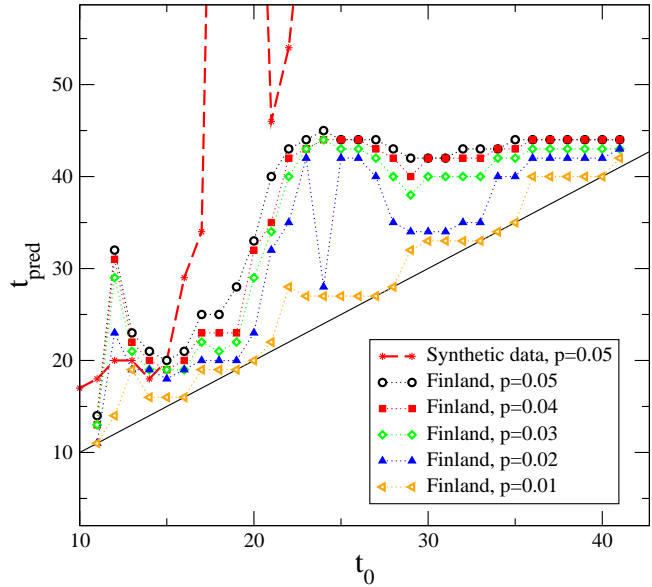}}
\caption{End time $t_{\rm pred}$  of correct predictions of syntheti	c data (stars) and Finland's real GDP  as a function of in-sample size $t_0$ for various tolerance parameter $p$. Dashed and dotted lines are for eye guidance only.\label{fig:Finland_tpred}}
\end{figure}

\begin{figure}
\label{fig:Finland_GDP}
\centerline{\includegraphics*[width=0.7\textwidth]{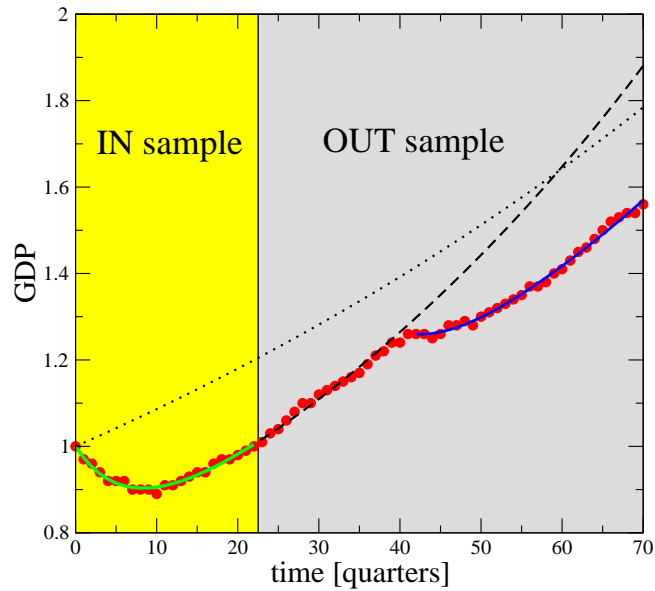}}
\caption{Finland's real GDP as a function of time, starting in October 1990. Red circles: quarterly data, green line: in-sample fit, black dashed line : out-sample prediction, blue line: fit starting from the 2001 slowdown, black dotted line: continuation of the pre-1990 GDP trend.}
\end{figure}

Choosing parameters that loosely replicate the evolution of the quarterly real GDP of Finland ($f=0.75$, $\lambda_+=0.0125$, $\lambda_-=-0.169$), and $\nu=0.005$, we generate a single 200-time steps time series (50 years). Taking an in-sample size of $t_0\in\{5,\dots,200\}$, we then find in the out-sample $t_{\rm pred}>t_0$ such that the relative difference between the prediction differs and the data at $t_{\rm pred}$ exceeds the tolerance $p=0.01, \cdots,0.05$. Even if $t_{\rm pred}$ as a function of $t_0$ depends on each realisation of the noise, a general pattern arises: at around $t_0=15-17$, which includes a part of the recovery, the prediction length increases tremendously, although at first with very large fluctuations whose details depend on the noise realisation, and reaches quickly the total length of the generated time series.

\subsection{Detecting additional shocks}
\label{subsec:Finland}
Armed with this positive conclusion, we apply the same procedure to the real quarterly, seasonally-adjusted, GDP of Finland. Fig.\ \ref{fig:Finland_tpred} clearly shows a saturation of $t_{\rm pred}\simeq 42$ for $t_0\ge 23$ even when the tolerance $p$ increases (at which point one can predict the GDP five years ahead) whereas the length of the time series is 71. This means that something happens at $t=42$, and indeed, it corresponds to the shock of the end of the Internet bubble. By decreasing the analysis sensitivity to fluctuations $p$, additional shocks can be detected. This is clear for $p=0.01$: additional plateaux at $t=27$ and $34$ can be seen.  Where to stop for $p$ is somewhat problematic for now and will be addressed in a future work.

\subsection{Shape rather than causes}
It is enlightening to look at a few well-known papers in Economics on the transition from communism to capitalism, e.g. \cite{FisherSahay,Sachs96,Popov2006}, which review the numerous aspects and precepts of economic transition and growth, together with various methodologies. On the one hand they are helpful reminders that an economy is an intricate system with many parameters of unknown influence, each of which being worth discussing, hence, that the parameters in our model are global measures of many processes of different types. On the other hand it is curious to note that, if most of them did plot GDP versus time, to our knowledge nobody tried to fit the time series of the GDP, which is all the more surprising since an equation similar {\em in fine} to Eq.\ \req{eq:g(t)} (but derived from other hypotheses) is found in an appendix of \cite{Popov2006}.

\begin{figure}
\centerline{\includegraphics[width=0.7\textwidth]{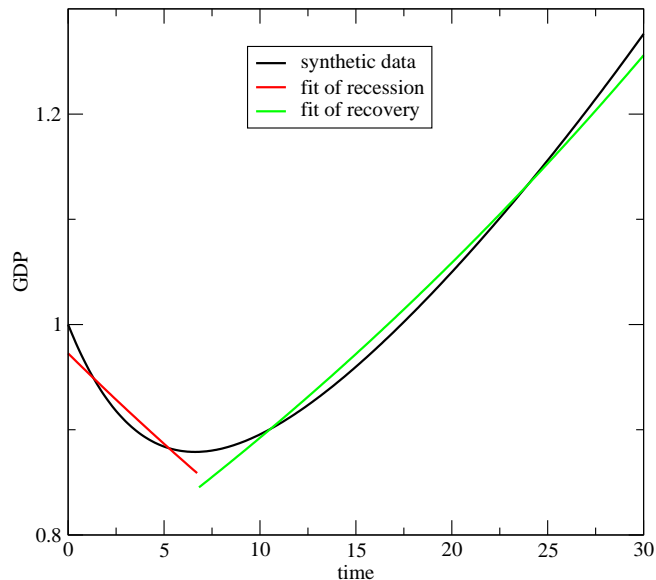}}
\caption{Synthetic GDP timeseries and two best exponential fits of the recession and recovery parts.}
\label{fig:synth}\end{figure}

A major difference between the approach of the respected economists cited above and the present fitting function is that
we focus on the typical temporal pattern or shape of recession and recovery after a shock (or the second one, when available) whereas, very understandably, the economists are concerned with finding the {\em causes} of recession and recovery and in particular of the variations between countries. Some means to assess the importance of reforms mostly come from abstract discussions on political economy. For instance, \cite{FisherSahay} singles out Uzbekistan and Belarus, noting the apparent contradiction between their economic performance and lack of reforms, nevertheless concluding that ``[...] it is reasonable to predict that they will grow more slowly than those who have undertaken more extensive reforms.'' However, our fits indicate that their parameters are in line with those of Lithuania in current dollars and Latvia in constant dollars.

One of the other tools widely used is factor analysis (known as multivariate linear fit in other disciplines) of the average growth in the recession and recovery parts (separately), where potentially relevant variables are guessed and then deemed actually relevant or not by statistical tests, yielding a regression coefficient $R^2\simeq0.5$; the major problem of this approach is that the decline and growth rates are evaluated on curves that are not exponentials but a superposition of several of them. Thus, neglecting the real curvature of the GDP gives biased estimates of both the decline and recovery rates, as illustrated in Fig.\ \ref{fig:synth}, which may explain in part the poor $R^2$ obtained by factor analysis. But it does make sense to apply this method to the fitted parameters ($f$, $\lambda_+$, $\lambda_-$)  in order to determine why the economic initial cer onditions were so different. It is worth to remark that $f$ does provide an important additional piece of information about the economy. We leave it to forthcoming publications.

Since our model is the simplest one to respect the underlying economic processes, it may not be entirely faithful, but its definition should not be over-interpreted. For instance, its success only says that the economy behaves very similarly to a two-sector approximation. It does not imply that the economy can be split into two parts inside which growth rates are equal (see also section \ref{sec:whyworks}).

\section{A simple theoretical model of economic activity transfer}
\label{sec:th}
The model used to fit data in the previous section can be derived from a simplification of the so-called AB model \citep{SolomonAB,SolomonAB_PRE}, a reaction-diffusion lattice model where discrete particles of two types diffuse, meet, reproduce and die auto-catalytically.  \cite{GurSolomonPoland} applied it recently to explain the temporal and spatial dynamics of economic growth in Poland. Whereas the latter work divided Poland into $N$ interacting geographical parts with diffusing elementary economic units, the present contribution is to show that the same model with $N=2$ is able to explain the temporal dynamics of many countries in difficult times. This allows us to restrict the number of parameters, hence, to explore more easily the dynamical properties of such models. It should also be noted that a sector can be either geographical, industrial, or abstract.

The rationale behind this model is the following. The after-shock economy is supposed to consist in two sectors, one with activity $w_1$, growing intrinsically at rate $\alpha_1>0$, and the other one with activity $w_2$ but intrinsically shrinking ($\alpha_2<0$). They interact through economic activity transfer taking place at rate $\beta$, according to the difference of activity. Mathematically,
\begin{eqnarray}
  \frac{\partial w_1 (t)}{\partial t} = \alpha_1  w_1 (t) + \beta [\avg{w(t)}
   - w_1 (t)] &  & \\
  \frac{\partial w_2 (t)}{\partial t} = \alpha_2 w_2 (t) + \beta [\avg{w(t)} - w_2 (t)], &  &
\end{eqnarray}
where $\avg{w}=(w_1+w_2)/2$.

The actual  result of the government's and investor's various policies is assumed to be equivalent to taking a fraction $\beta/2$ of the difference of activity  between the two sectors from the largest sector and giving it to the smallest one. In other words, $\beta$ may not be the transfer rate wished for by the government, but the actual transfer rate (the distinction is valid for the rest of the discussion). This means in particular that, when the intrinsically expanding sector represents a small part of the economy, resources are transferred to it from the shrinking sector, thereby accelerating the transition. Note that because of redistribution, both sectors end up growing at the same rate. Therefore, it is wrong to think of the dynamics of this model as describing a growing sector and a declining sector since both have a growing and a declining part.

Solving the dynamics of this system is straightforward by computing the eigenvalues and associated eigenvectors of the above coupled dynamical equations, following standard procedure. The two eigenvalues are
\be
\lambda_{\pm}=\frac{\delta[\sigma/\delta-\zeta\pm\sqrt{1+\zeta^2}]}{2}
\ee
where $\delta=\alpha_1-\alpha_2$, $\sigma=\alpha_1+\alpha_2$ and $\zeta=\beta/\delta$. These eigenvalues correspond to the rates measured in the previous section.
 The unnormalised eigenvectors are $(\zeta,-1\pm\sqrt{1+\zeta^2})$. Let us denote by $\mathbf{v_\pm}=(v_{\pm.1},v_{\pm_2})$ the respective orthonormal eigenvectors.
Following standard procedure, one decomposes $\mathbf{w}(t=0)$ into the basis $\mathbf{v_\pm}$, obtaining $\mathbf{w}(t)=\omega_+\mathbf{v_+}e^{\lambda_+t}+\omega_-\mathbf{v_-}e^{\lambda_-t}$ where $\omega_\pm=\mathbf{w(0)}.\mathbf{v}_\pm$ are the projections of the initial conditions onto the sector decomposition described above. In other words, both $w_1$ and $w_2$ have an increasing and a decreasing part.  The steady state is reached when the importance of the negative component is vanishingly small compared to the positive component both for $w_1$ and $w_2$. The typical time for reaching this asymptotic regime is $O(1/(\lambda_++|\lambda_-|))$ units of time. Then the two groups grow at the same rate, $\lambda_+$ (Fig.\ \ref{fig:w1w2}). In this regime, the growth of the negative component is entirely due to the transfer of economic activity from the positive component.

We shall be interested in this paper in the total economic activity $W=w_1+w_2$ and shall consider the GDP as its proxy. Also we are interested in the dynamics of inequality between the sectors, measured by $\Delta=w_1/w_2$. Note that the empirical data determine only partially the parameters of Eq.\ \req{eq:g(t)} of previous section: while the rates $\lambda_\pm$ can be measured directly, more detailed information is needed in order to determine all three parameters $\alpha_1$, $\alpha_2$ and $\beta$. This is due to the fact that $f$ does not correspond directly to $w_1(0)$ since even at the beginning sector $2$ has a growing part (i.e. $v_2^+\ne 0$).

\subsection{Why this model works}
\label{sec:whyworks}
The contrast between the intricacies of economic policy making and implementation \citep{FisherSahay} and the simplicity of our model on the one hand, and the quality of our fits and the (relatively) poor explanatory power of factor-based growth analysis \citep{FisherSahay,Popov2006} on the other hand is perplexing. In order to understand the surprisingly good performance of the simple fitting function of Eq.\ \req{eq:g(t)}, one needs to go back to the two-dimensional autocatalytic AB model of \cite{SolomonAB}, which describes spatially-distributed and heterogeneous logistic systems. Its ability to reproduce both the spatial and temporal dynamics of Poland's GDP is striking \citep{GurSolomonPoland}; interestingly, it finds that the local level of education is the most relevant factor in growth, in line with \cite{FisherSahay}.  In the case of Poland, it predicted successfully the pattern of recession and recovery of the parts of Poland: the activity of each part of the country reaches its minimum at different times, while the final growth rate is the same for all parts, strongly suggesting that an economic activity transfer process is at work; in other words, plotting the economic activity evolution of various sectors as a function of time produces a whole variety of J-shaped time-series, all ending up with the same growth rate. The simplification to two parts, or two sectors, works because the economy is an autocatalytic, that is, multiplicative,  process: the parts that reach their minima later than the best part often have shrunk so much that their contribution to the total GDP is negligible afterwards.  The use of quarterly data for Finland and the UK, as shown above, seems to indicate that the fitting equation we propose describes very well the GDP.

\subsection{Static policy making}

Assume that the rate $\alpha_1$ and $\alpha_2$ are constant and fixed by constraints beyond the control of the government. The government's only influence is in the transfer rate through the tax rate policy. This in itself is a very powerful instrument that the government is pressed to use:  indeed economic activity is linked to employment, and a fast-shrinking sector implies growing social inequality and voters dissatisfaction. If the rate of shrinking is much faster than the rate of labour transfer between the two sectors, inequality at the sector level translates into growing social inequality. In that sense, the inequality between sectors is an upper bound to social inequality. It should be noted that $\beta$ is the effective rate of transfer, not the one hoped for by the government; indeed, if the latter is not able to collect taxes or if its authority is undermined by inadequate rule of law due to the collapse of institutions, the effective $\beta$ may turn out much smaller.

The final growth rate depends much on the policy: increasing $\beta$ reduces both eigenvalues, hence the total growth rate {\em in the steady state}: maximal asymptotic economic growth is achieved when there are no transfer of wealth. This seems to substantiate the claims of the so-called supply-side economics (see e.g. \cite{Lucas-supply_side}). However, global growth rate is not the only success measure of a taxation policy: inequality is also to be taken into account.

Indeed,  since $\lambda_-<0$, group $2$ would simply disappear in the absence of redistribution. Decision makers who only focus on growth will therefore take $\beta$ as small as socially responsible and electorally possible. Some others will try to minimise inequality between sectors. Since both $w_1$ and $w_2$ end up growing at the same speed, their asymptotic ratio $\Delta=\lim_{t\to\infty}w_1(t)/w_2(t)$ is a measure of economic inequality. Using basic algebra, one finds that
\be
\Delta=v_{+,1}/v_{+,2}=1/(-1/\zeta+\sqrt{1+\zeta^2})\simeq 2/\zeta=2(\alpha_1-\alpha_2)/\beta
\ee
 if $\zeta\ll1$, in which case reducing the sector inequality by a half requires to double the transfer rate; in addition, sector inequality is proportional to the difference of growth rate.

Since the rates are fixed by assumption, sector inequality only depends on the transfer rate, not on initial conditions. Inequality ceases to exist only for large $\beta$, at the cost of growth rate.

\begin{figure}
 \centerline{\includegraphics*[width=0.7\textwidth]{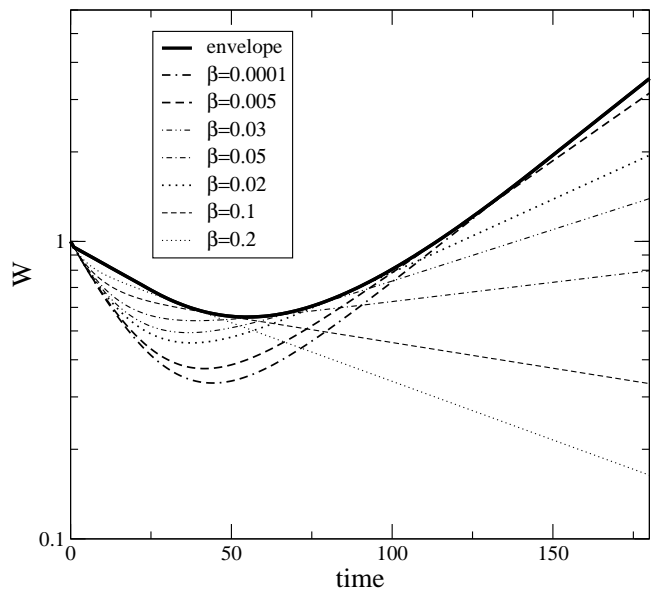}}
\caption{Total economic output versus time for various values of the transfer rate $\beta$ ($\alpha_1=0.02$, $\alpha_2=-0.05$, $w_1(0)=0.1$, $w_2(0)=0.9$), and the (virtual) upper envelope of all  curves (thick black line).}
\label{fig:envelope}
\end{figure}

Therefore, assuming fixed transfer rate, a head of state of a country facing a recession may be able to choose between a small but long recession with anemic final growth, or a large but short-lasting recession with large final growth (Fig. \ref{fig:envelope}). A cynical politician would ensure that the wealth of the majority of voters has increased by the end of his tenure or at least that the recovery has begun.

\subsection{Gradualism versus shock therapy}

All the Eastern European countries have experienced economic recession when switching from communism to capitalism. The variety of intrinsic growth and decline rates and policies yielded vast differences between speed of recovery and depth of recession of various countries. Understandably a large corpus of literature investigates what factors could explain this variety of behaviours \citep{Sachs96,FisherSahay,Popov2006,ULJ}. In particular, the technique of making the transition abrupt and short has been labelled as shock therapy \citep{Kolodko}. The concept of shock therapy has been the focus on long debates which have not been settled to this very day  \citep{Lucas-supply_side}. The other approach is called gradualism, and advocates to follow a more gentle rhythm \citep{Kolodko}.

Our model makes it possible to investigate this issue. We shall assume that $\alpha_1$ and $\alpha_2$ are intrinsic to the economy and therefore constant; the government influences the economy by trying to adjust the effective transfer rate $\beta(t)$. The shock therapy consists in lowering abruptly $\beta$ from the high level of communism to the small level of capitalism. Gradualism implies a smoother mathematical function for $\beta(t)$.

\subsubsection{Constant policy: shock therapy}

Figure \ref{fig:envelope} plots various scenarios for $W(t)$ at constant $\beta$ and shows the influence of $\beta$ on the outcome. The cases with small $\beta$ correspond to shock therapy. They are characterised by a deep recession and both a faster final growth rate and accordingly a higher GDP. Therefore, after many years, the tenants of this policy are vindicated since their courageous but harsh recommendations are proved correct as regards the growth rate of GDP and value {\em compared} to other static policies. This view is right, but only in a static context, as it maximises the {\em final} growth rate, not the instantaneous one, therefore not the actual GDP (see below),  and deep recession ensues.

\subsubsection{Dynamical policy: envelope}

\begin{figure}
 \centerline{\includegraphics*[width=0.7\textwidth]{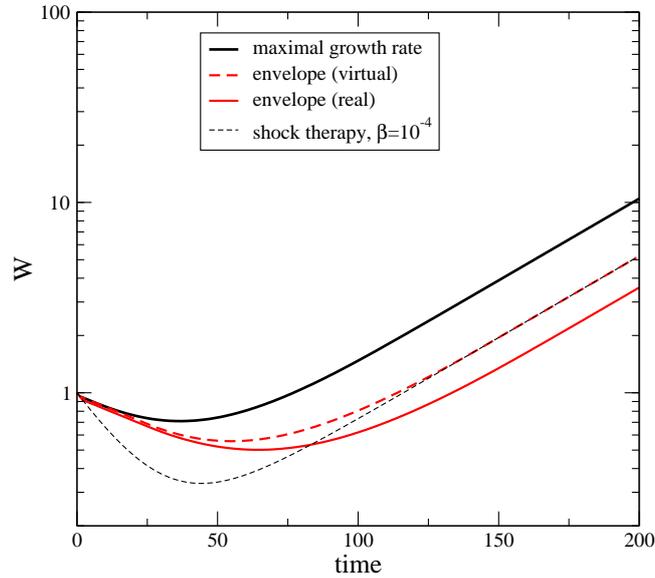}}
\caption{Total economic output $W$ for the optimal policy (black line), the envelope-based gradual policy (red lines) and the shock therapy (dashed black line). Same parameters as in Fig 2.}
\label{fig:envelope_bopt}
\end{figure}

Few experiences are more frustrating for a politician than to have implemented a policy that will lead to the recovery of one's country, but too late to be re-elected. Instead of heroically jeopardising one's political career, one should ask how to implement a policy that would avoid most troubles.

There is another way of looking at this figure: what if one could stay on the upper envelope of all the scenarios and thereby avoiding as much as possible difficult times while maximising the final growth rate? This clearly needs a dynamical policy, i.e., $\beta(t)$. Running a thousand scenarios $W_i(t)$ with various $\beta_i$,  and selecting at each time $t$ the value of $\beta$ that corresponds to the maximal $W$ yields $\beta_{\rm env}(t)= \beta_{\arg \max_i W_i(t)} $ shown in Fig.\ \ref{fig:envelope}: redistribution should be kept maximal for a while, then $\beta_{\rm env}$ decreases exponentially fast in the region encompassing the worst phase of the recession,  and then decreases faster than exponentially. Regretfully for a head of state, this view is purely virtual: Figure\ \ref{fig:envelope_bopt} reports the GDP $W[\beta_{\rm env}]$ actually obtained by using the policy $\beta_{\rm env}(t)$, the envelope $\max_i W_i(t)$ of Fig.\ 2 and an example of shock therapy with constant $\beta$ (forget  $W_{\rm opt}$ for the time being). The difference  between the virtual and real $W$ come from the fact that it is impossible to stay on the envelope by controlling $\beta(t)$ on the basis of $W$ alone: when $W(t)$s of two scenarios cross, their components $w_1(t)$ and $w_2(t)$ are not equal. Thus shock therapy works better than trying to stay on the envelope.

Curiously,  $W(t)$ actually obtainable by using this method is also relatively well fitted with Eq.\ \ref{eq:g(t)}, i.e. with constant parameters, and gives for the curve reported in Fig.\ \ref{fig:envelope} $\lambda_+\simeq0.020$, $\lambda_-\simeq-0.019$, and $f=0.068$ with uncertainties smaller than a percent, whereas the initial values were 0.02, -0.05 and 0.1, respectively. In other words, the effective shrinking rate is reduced, the rate of growth of the expanding sector remains unchanged (which is needed in order to retain the same final growth rate), while the apparent fraction of the expanding sector decreases considerably; interestingly, the difference between the fits of the envelope itself and the attainable $W(t)$ is limited to $f$: the envelope has the same apparent $f$ as the individual runs.

\subsubsection{Optimal policy: maximal $W$}

Another policy is to maximising $W$ at each time step, which reduces to the maximisation of the growth rate with respect to $\beta$: $\frac{\partial \dot W}{\partial \beta}=0$. This leads to a transcendental equation to be solved numerically at each time step. The resulting $W_{\beta_{\rm opt}}$ is reported in Fig.\ \ref{fig:envelope_bopt}, which shows unambiguously the benefits of the proposed optimal dynamical policy, that is, of gradualism with respect to shock therapy. Indeed, the value of the GDP in the recovery phase is increased several times with respect to static policies and with respect to envelope-based dynamical policies, while sharing the same asymptotic growth rate. We therefore claim that shock therapies are unadapted to economies in crises as regards GDP. Patience and gradualism are better solutions in this kind of situations.

\begin{figure}
 \centerline{\includegraphics*[width=0.7\textwidth]{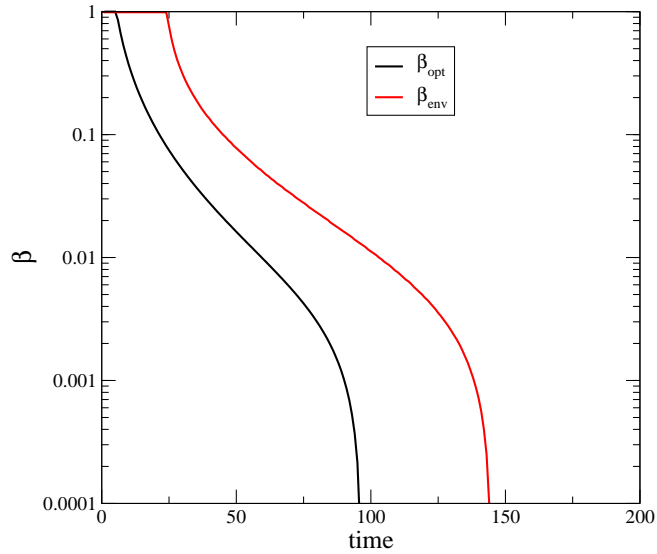}}
\caption{
Optimal value of the transfer rate $\beta$ versus time for the  envelope-based policy (red line) and the optimal policy (black line). Same parameters as in Fig.\ \ref{fig:envelope}.}
\label{fig:bopt}
\end{figure}

Looking at the optimal value of $\beta$ (Fig.\ \ref{fig:bopt}) reveals that indeed taxes should decrease rapidly, but not instantaneously. This means that the intuition behind shock therapies is correct, but only in the later stages of the time evolution. What matters is the road to minimum taxes, all the more since the economy follows multiplicative processes: optimising it may change tremendously the fate of countries and people, as shown by the results of envelope-based and optimal policies.

Fitting $W_{\beta_{\rm opt}}$ with Eq.\ \req{eq:g(t)} yields $f\simeq0.80$, $\lambda_+\simeq0.20$ and $\lambda_-\simeq-0.027$. Therefore, the optimal policy both increases the apparent fraction of the growing part of the economy and decreases the apparent rate of shrinking of the decaying sector, while of course keeping constant the final rate of growth.

\subsubsection{Detecting static, envelope-based, and optimal policies}

\begin{figure}
 \centerline{\includegraphics*[width=0.7\textwidth]{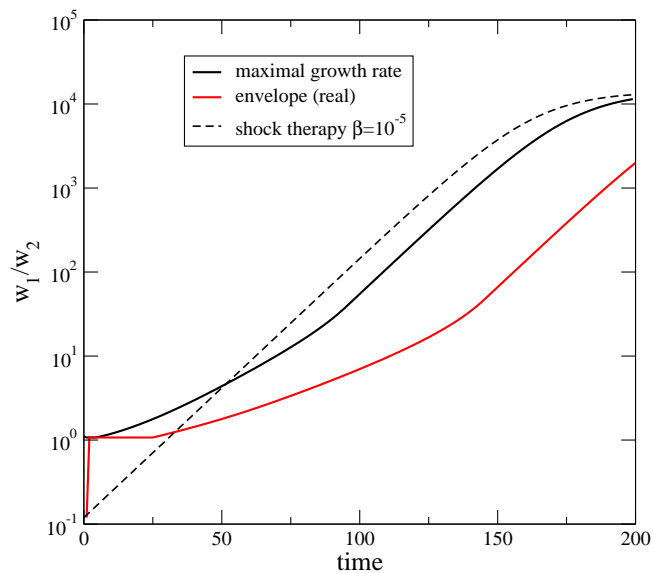}}
\caption{Sector inequality as a function of time for the optimal policy (black line), the envelope-based gradual policy (red line) and the shock therapy (dashed black line). Same parameters as in Fig 2.}
\label{fig:w1w2}
\end{figure}

A somewhat frustrating result of the previous two dynamical policies is the impossibility to distinguish them from a static one. Indeed,  in the absence of additional information about the applied economic policies, one cannot reconstruct it from the GDP time series. For that purpose, one would need data about at least two sectors. Plotting $\Delta=w_1/w_2$ as function of time allows one to distinguish a static, envelope-based and optimal policy, as reported in Fig.\ \ref{fig:w1w2}: a static policy has a negative curvature, while dynamic ones start with a flat line, followed with a positive curvature and then an inflexion point. All of them reach the same asymptotic values since one imposed a minimum $\beta=0.00001$ in order to compare the three policies. It may be difficult in practice to discriminate with the naked eye an envelope-based policy from the optimal one, but easy to detect a static policy \footnote{It is clear that increasing taxes will also lead to a negative curvature; however, we assume that the transition from communism to capitalism needs a decrease of taxes.}. However, in further investigations one could measure $w_1$ and $w_2$, and thus determine all the parameters of the model, including $\beta_{\rm opt}(t)$.

\section{Discussion}

The idea that the dynamics of economies is affected by shocks is by no means new (see e.g. \cite{Slutzky}). The contribution of the present work is to show that a rare single negative shock, well defined in time, has a lasting influence on the GDP evolution because of the existence of an intrinsic response function and that it is sufficient to lead to a recession.

We claim that the simple function of Eq \req{eq:g(t)} is at the very least a very good approximation to the response function: it fits well the GDP of countries in various continents, at various levels of industrialisation, at different times, short and long recessions after large shocks, and  even small bumps after small shocks. This is indeed what can be expected from a response function. In addition, it suggests that the GDP time series are quite smooth between infrequent shocks.

It may be argued that relying on shocks to justify deviations of the GDP from the fitting equation is bad practice, hence, that splitting time series into segments, each of them obeying the fitting function amounts to relegate some part of the problem to external causes. This would be the case if the fitting function could not work for more than a few points, i.e. if the pseudo-shocks were relatively frequent. In addition, the fact that the fits are even more impressive for quarterly data is a strong argument in favour of our equation. It remains that the shocks themselves deserve a more thorough study, which is left to future work. In particular, one has to find a statistical criterion for the splitting of a time series into segments, and to address the problem of shocks that propagate gradually to a given economy, yielding a concave GDP at the start of recessions, possibly needing to consider a dynamical fraction $f$.

Although we gave intuitive and simple theoretical arguments to motivate the mathematical shape of this function, one ideally wishes also to build a model whose response function to external shocks is compatible with our findings. We are not aware of any.

Finally, our model suggests that lasting recessions are smooth curves when viewed at quarterly data level. This means in particular that they have a well-marked minimum corresponding to several quarters of sluggish growth; therefore, hopes to avoid that part of the curve are misguided. It is hard to make predictions of the GDP evolution regarding current recessions first because there are only three or four quarterly data estimates, which are often revised in a sizeable proportion in times of recessions, and also because currently available quarterly data does not show any unambibuous sign of inflexion.

\bibliographystyle{econ}
\bibliography{biblio}

\newpage
\appendix
\section{Model fitting}

In the following tables, {\sc srs} stands for square residuals sum, {\sc srm} for  square residuals mean,  {\sc rsrs} for relative square residuals sum,  {\sc rsrm} for relative square residuals mean;  years indicates the number of years along which the fit has been done. The table contains a wealth of data which can be further exploited and whose potential has been only scratched by the present publication.

\begin{center}
\begin{table}[h!p!b! width=10cm]
\begin{tabular}{|l||c|c|c|c|c|c|c|c|c|c|c|}
\hline Country & first year & last year & $f$ & $e^{\lambda_+}$ &
$e^{\lambda_-}$ & {\sc srs} & {\sc srm} &{\sc  rsrs} & {\sc rsrm }& years\\
\hline Albania & 1990 & 2006 & 59.5 & 1.064 & 0.367 & 301.97 &
17.763 & 402.92 & 23.701 & 17 \\
\hline Bahamas & 1990 & 1999 & 70.8 & 1.061 & 0.662 & 4.6524 &
0.4652 & 4.4493 & 0.4449 & 10 \\
\hline Belarus & 1991 & 2006 & 38.0 & 1.089 & 0.737 & 151.97 &
9.4982 & 241.12 & 15.070 & 16 \\
\hline Bolivia & 1981 & 1998 & 61.9 & 1.049 & 0.775 & 16.529 &
0.9183 & 18.257 & 1.0143 & 18 \\
\hline Chile & 1981 & 1997 & 51.2 & 1.079 & 0.522 & 56.499 & 3.3235
& 37.112 & 2.1830 & 17 \\
\hline Estonia & 1990 & 2006 & 40.7 & 1.083 & 0.688 & 148.97 &
8.7630 & 211.69 & 12.452 & 17 \\
\hline Finland & 1990 & 2006 & 82.1 & 1.036 & 0.367 & 79.249 &
4.6617 & 65.940 & 3.8788 & 17 \\
\hline Hungary & 1989 & 2006 & 64.0 & 1.046 & 0.694 & 38.588 &
2.1438 & 45.955 & 2.5530 & 18 \\
\hline Kazakhstan & 1990 & 2006 & 84.4 & 0.837 & 1.138 & 105.23 &
6.1903 & 186.39 & 10.964 & 17 \\
\hline Latvia & 1990 & 2006 & 27.4 & 1.087 & 0.651 & 245.92 & 14.466
& 508.97 & 29.939 & 17 \\
\hline Lithuania & 1990 & 2006 & 31.1 & 1.082 & 0.723 & 214.11 &
12.594 & 411.24 & 24.191 & 17 \\
\hline Mongolia & 1990 & 2005 & 64.5 & 1.045 & 0.615 & 156.65 &
9.7910 & 143.21 & 8.9512 & 16 \\
\hline Moldova & 1990 & 2006 & 14.6 & 1.081 & 0.734 &
103.41 & 6.0834 & 395.19 & 23.246 & 17 \\
\hline Russia& 1990 & 2006 & 20.4 & 1.100 & 0.825 &
102.63 & 6.0373 & 181.19 & 10.658 & 17 \\
\hline Slovakia & 1990 & 2006 & 72.9 & 1.042 & 0.367 & 102.28 &
6.0170 & 95.323 & 5.6072 & 17 \\
\hline Slovenia & 1990 & 2006 & 79.5 & 1.040 & 0.367 & 29.452 &
1.7324 & 27.235 & 1.6020 & 17 \\
\hline Sweden & 1990 & 2006 & 88.1 & 1.029 & 0.549 & 28.900 & 1.7000
& 23.565 & 1.3861 & 17 \\
\hline Turkmenistan & 1990 & 2006 & 29.9 & 1.080 & 0.809 & 291.63 &
17.154 & 538.89 & 31.699 & 17 \\
\hline Ukraine & 1990 & 2006 & 6.71 & 1.153 & 0.835 & 241.22 &
14.189 & 656.64 & 38.626 & 17 \\
\hline Uzbekistan & 1990 & 2006 & 42.6 & 1.074 & 0.829 & 77.763 &
4.5743 & 87.654 & 5.1561 & 17 \\
\hline Bulgaria & 1988 & 1995 & 0.17 & 1.967 & 0.929 & 22.261 &
2.7827 & 30.514 & 3.8143 & 8 \\
\hline Bulgaria & 1995 & 2005 & 64.5 & 1.051 & 0.463 & 5.0247 &
0.4567 & 6.0662 & 0.5514 & 11 \\
\hline Czechia & 1990 & 1996 & 75.0 & 1.047 & 0.422 & 3.3792
& 0.4827 & 4.2602 & 0.6086 & 7 \\
\hline Czechia& 1996 & 2006 & 38.1 & 1.102 & 0.927 & 5.8393
& 0.5308 & 5.3579 & 0.4870 & 11 \\
\hline Romania & 1990 & 1996 & 60.4 & 1.075 & 0.526 & 6.2065 &
0.8866 & 8.4506 & 1.2072 & 7 \\
\hline Romania & 1996 & 2006 & 60.3 & 1.070 & 0.656 & 9.3063 &
0.8460 & 9.1387 & 0.8307 & 11 \\
\hline
\end{tabular}
\caption{GDP in current dollars}
\end{table}
\end{center}

\begin{center}
\begin{table}[h!p!b! width=10cm]
\begin{tabular}{|l||c|c|c|c|c|c|c|c|c|c|c|}
\hline Country & first year & last year & $f$ & $e^{\lambda_+}$ &
$e^{\lambda_-}$ & {\sc srs} & {\sc srm} &{\sc  rsrs} & {\sc rsrm }& years\\
\hline Albania & 1989 & 2004 & 51.0 & 1.064 & 0.514 & 341.56 &
21.347 & 567.59 & 35.474 & 16 \\
\hline Bahamas & 1990 & 2000 & 52.8 & 1.075 & 0.863 & 4.4978 &
0.4088 & 4.5325 & 0.4120 & 11 \\
\hline Belarus & 1990 & 2004 & 26.6 & 1.106 & 0.834 & 278.38 &
18.559 & 437.38 & 29.158 & 15 \\
\hline Bolivia & 1981 & 1998 & 61.0 & 1.050 & 0.793 & 19.853 &
1.1029 & 21.966 & 1.2203 & 18 \\
\hline Chile & 1981 & 1998 & 73.1 & 1.076 & 0.367 & 167.27 & 9.2928
& 51.868 & 2.8815 & 18 \\
\hline Estonia & 1989 & 2004 & 35.2 & 1.080 & 0.748 & 204.50 &
12.781 & 362.29 & 22.643 & 16 \\
\hline Finland & 1990 & 2004 & 81.2 & 1.036 & 0.392 & 61.378 &
4.0919 & 53.024 & 3.5349 & 15 \\
\hline Hungary & 1989 & 2004 & 60.9 & 1.048 & 0.701 & 41.342 &
2.5838 & 48.124 & 3.0077 & 16 \\
\hline Kazakhstan & 1990 & 2004 & 10.5 & 1.170 & 0.857 & 71.987 &
4.7991 & 129.56 & 8.6377 & 15 \\
\hline Latvia & 1989 & 2004 & 24.9 & 1.090 & 0.740 & 305.80 & 19.113
& 655.52 & 40.970 & 16 \\
\hline Lithuania & 1990 & 1998 & 4.42 & 1.352 & 0.830 & 122.13 &
13.571 & 217.51 & 24.167 & 9 \\
\hline Lithuania & 1998 & 2004 & 84.9 & 1.081 & 0.422 & 2.3325 &
0.3332 & 1.5147 & 0.2163 & 7 \\
\hline Mongolia & 1990 & 1995 & 0.31 & 2.701 & 0.899 & 1.6940 &
0.2823 & 2.5702 & 0.4283 & 6 \\
\hline Mongolia & 1995 & 2004 & 11.7 & 1.173 & 1 & 64.007 & 6.4007 &
45.001 & 4.5001 & 10 \\
\hline Moldova & 1989 & 2004 & 2.56 & 1.200 & 0.843 & 359.69 &
22.480 & 913.24 & 57.078 & 16 \\
\hline Russia& 1989 & 2004 & 5.19 & 1.185 & 0.895 &
191.79 & 11.987 & 341.91 & 21.369 & 16 \\
\hline Slovakia & 1989 & 1998 & 37.1 & 1.106 & 0.778 & 91.756 &
9.1756 & 122.02 & 12.202 & 10 \\
\hline Slovakia & 1998 & 2004 & 83.5 & 1.063 & 0.753 & 0.2443 &
0.0349 & 0.2118 & 0.0302 & 7 \\
\hline Slovenia & 1990 & 2004 & 79.4 & 1.039 & 0.367 & 25.088 &
1.6725 & 23.178 & 1.5452 & 15 \\
\hline Sweden & 1990 & 2004 & 89.0 & 1.029 & 0.508 & 32.082 & 2.1388
& 26.378 & 1.7585 & 15 \\
\hline Turkmenistan & 1990 & 2001 & 0.12 & 1.744 & 0.908 & 183.33 &
15.277 & 359.04 & 29.920 & 12 \\
\hline Ukraine & 1989 & 2004 & 0.61 & 1.330 & 0.882 & 274.88 &
17.180 & 812.32 & 50.770 & 16 \\
\hline Uzbekistan & 1990 & 2004 & 49.1 & 1.063 & 0.802 & 69.781 &
4.6520 & 83.878 & 5.5918 & 15 \\
\hline Bulgaria & 1988 & 1995 & 0.15 & 1.968 & 0.929 & 22.327 &
2.7909 & 30.615 & 3.8269 & 8 \\
\hline Bulgaria & 1995 & 2004 & 74.1 & 1.050 & 0.452 & 4.6503 &
0.4650 & 5.8248 & 0.5824 & 10 \\
\hline Czechia& 1990 & 1996 & 74.8 & 1.047 & 0.427 & 3.3822 &
0.4831 & 4.2688 & 0.6098 & 7 \\
\hline Czechia& 1996 & 2004 & 86.9 & 1.035 & 0.649 & 4.3757 &
0.4861 & 4.0627 & 0.4514 & 9 \\
\hline Romania & 1988 & 1996 & 2.67 & 1.411 & 0.901 & 61.438 &
6.8265 & 95.846 & 10.649 & 9 \\
\hline Romania & 1996 & 2004 & 55.3 & 1.090 & 0.729 & 3.6354 &
0.4039 & 4.0334 & 0.4481 & 9 \\
\hline
\end{tabular}
\caption{GDP in constant dollars}
\end{table}
\end{center}

\end{document}